\documentclass[prb,preprint,aps,showpacs,longbibliography,
lengthcheck,superscriptaddress]{revtex4-1}

\usepackage{graphicx}%
\usepackage{dcolumn}
\usepackage{float}

\newcommand{\sinp}{Saha Institute of Nuclear Physics, I/AF Bidhannagar, Kolkata 700 064, India}
\newcommand{\scott}{Department of Physics, Scottish Church College, 1 $\&$ 3 Urquhart Square, Kolkata 700 006, India}

\begin{document}

\title{Scaling of NonOhmic Conduction in Strongly Correlated Systems}

\author{D. Talukdar}
\affiliation{\sinp}

\author{U. N. Nandi}%
\affiliation{\scott}
	
\author{A. Poddar}
\affiliation{\sinp}

\author{P. Mandal}%
\affiliation{\sinp}

\author{K. K. Bardhan}%
\affiliation{\sinp}
\email{kamalk.bardhan@saha.ac.in}

\date{\today}

\begin{abstract}
A new scaling formalism is used to analyze nonlinear \textit{I-V} data in the vicinity of metal-insulator transitions (MIT) in five manganite systems. An exponent, called nonlinearity exponent, and an onset field for nonlinearity, both characteristic of the system under study, are obtained from the analysis. The onset field is found to have an anomalously low value corroborating the theoretically predicted electronically soft phases. The scaling functions above and below the MIT of a polycrystalline sample are found to be same but with different exponents which are attributed to the distribution of the MIT temperatures. The applicability of the scaling in manganites underlines the universal response of the disordered systems to electric field.
\end{abstract}

 
\maketitle

\section{Introduction}
The nonOhmic response to an applied electric field is quite common in disordered phases or systems which include recent systems of interest such as manganites, conducting polymers. It has now assumed critical importance because of increasing applied use of various low dimensional nanostructures. The conduction becomes nonOhmic even at small biases used in laboratories. In spite of considerable theoretical efforts\cite {ladieu00} spent in last several decades towards understanding of the mechanism of nonOhmic hopping transport, the latter still remains an open issue. It has been recently shown that systems exhibiting variable range hopping (VRH) in three dimension such as conducting polymers and amorphous semiconductors possess a single field scale leading to a simple, yet nontrivial scaling description\cite {talukdar11} of the field-dependent conductivity $\sigma (M,F)$:
\begin{equation}
{\sigma(M,F) \over \sigma(M,0)} = \Phi \left ({F \over F_o} \right ).
\label{eq:scaling}
\end{equation}
Here \textit{F} is the applied dc electric field, $\Phi$ is a scaling function, \textit{M} is a physical variable (e.g., temperature) or a group of variables. The field scale $F_o(M)$ is given by
\begin{equation}
F_o (M) \sim {\sigma_o}^{x_M},
\label{eq:fscale}
\end{equation}
where $\sigma_o(M)=\sigma(M,0)$ is the zero bias conductivity and $x_M$ is called the nonlinearity exponent. Eq. \ref {eq:fscale} is similar to the power-laws of critical phenomena in phase transitions. A scaling relation exhibits model-independent properties which are particularly useful in the absence of a definitive theory as in the present case. For example, the above relations when applied to VRH systems have led to several conclusions which are very different from those in the current theories \cite {talukdar11} such as the power-law, in contrast to the predicted exponential, dependence of conductivities at large fields. The conductance $\Sigma$ here is simply defined as the ratio $I/V$.

A key issue concerns scope of the scaling (Eq. \ref {eq:scaling}) i.e. whether such a scaling is a universal property of the disordered systems even in the presence of strong electronic correlation as in manganites. This provided the present motivation for studying manganites which are known to have rich and complex disordered phases. Disorder, aside from spin disorder, arises from random potential fluctuations due to trivalent rare-earth and divalent alkaline-earth ion cores, Jahn-Teller distortion and local trapping in ferromagnetic regions with non-collinear magnetic order\cite {Coey95}. Some parts of these phases are inhomogeneous - not chemically or structurally but electronically - in ways that are different from say, granular composites or amorphous semiconductors which are usually treated within one-electron formalism\cite {abeles75,shklov76}. In an intermediate doped manganite a metal-insulator transition (MIT) usually manifests itself in the form of a resistance peak at a temperature $T_{MI}$ close to the Curie temperature $T_{c}$. The regime near a MIT exhibits colossal magnetoresistance. The resistance curve is rather asymmetric with a sharp drop on the lower side of $T_{MI}$, especially in single crystals. Spatial microscopy experiments\cite {uehara99, zhang02} and theories\cite {yunoki98, dagottobook} attribute this to nanoscale phase separation near the transition where the spin-sensitive transport occurs through a percolative network of ferromagnetic metallic (FM) and paramagnetic insulating (PI) domains. The intrinsic inhomogeneity due to phase separation creates conditions favorable for nonlinear (or nonOhmic) transport due to wide variations of local electric fields inside a sample. Mosgnyaga et al.\cite {moshnyaga09} investigated the MIT by means of electric third harmonic resistance ($R_{3w}$) and found that the latter is drastically enhanced near the MIT. The electrical nonlinear behavior was argued to arise from coupling of correlated polarons to the electric field. Nonlinear conduction has been earlier studied in manganites mostly in charge ordered (CO) regimes\cite{guha00b, *pradhan04, *odagawa04}, also in materials without CO\cite {yuzhelevski01, *tanaka02, *mercone02}, and in orbital ordered samples\cite {mondal11}. However, the analysis of the nonlinear data so far has been devoid of any systematics. 

In this paper, we present evidence that states around a metal-insulator transition peak indeed obey the scaling (Eq. \ref{eq:scaling}) thus providing further evidence for universality of the scaling in disordered systems. It holds true in both single crystal and polycrystalline samples of manganites, thus proving that the origin of nonlinearity is intrinsic, and not due to extraneous factors such as inelastic intergrain tunneling. The scaling functions on both sides of a MIT in a polycrystalline sample are found to be identical albeit with different nonlinearity exponents $x_T$. The electric field scale $F_o$ is anomalously found to be of the order of 1-10 V/cm, about three orders of magnitude less than that in granular composites, corroborating the claim of electronically soft phases in manganites\cite {milward05}. A variety of such results on the other hand suggest that the field dependent transport in manganites can be used also as a probe of complexities in conduction mechanisms beyond those revealed in zero-bias measurements. It is shown further how the traditional method of probing nonlinear conduction by measurement of third harmonic voltage can be used as an alternative way to extract the nonlinearity exponent.

\begin{figure}
\includegraphics[width=5.8cm]{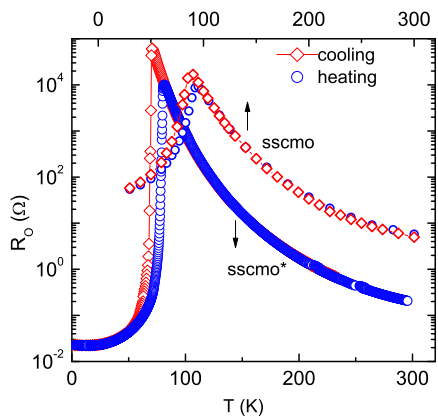}
\caption{(Color online) Ohmic resistance $R_o$ vs. temperature $T$ of two SSCMO samples - one single crystal (with star) and another polycrystalline. The symbols, circles and diamonds, correspond to heating and cooling cycles respectively. Peak temperatures $T_{MI}$ are 81 and 94 K respectively.}
\label{fig.1}
\end{figure}

\section{Experimental}
An important factor in choosing samples (first five in Table I) for the present study was the requirement to have resistance changes between the baseline and the peak on either side of a MIT as large as possible to extract reliably the nonlinearity exponent (see Eq. \ref {eq:fscale}). Accordingly, systems with Sm were chosen\cite {tokura06}. $\rm{La}_{0.275} \rm{Pr}_{0.35}\rm{Ca}_{0.375} \rm{MnO}_3$ was chosen for its robust phase separation property\cite {uehara99,podzorov01}. A single crystal of $\rm{Sm}_{0.55} (\rm{Sr}_{0.5} \rm{Ca}_{0.5})_{0.45}\rm{MnO}_3$ (SSCMO* in Table I) was grown using the floating zone technique\cite {sarkar09}. Four different polycrystalline samples (ones in Table I without star sign) were prepared by the usual solid-state reactions. Two SSCMO samples have slightly different chemical composition. Of all the samples, LMO is self-doped\cite {de05} while rest are the usual hole-doped. The values of $T_{MI}$ for different samples analyzed in this work are shown in Table I. \textit{{I-V}} measurements were done in a Janis cryotip with bar-shaped samples placed on sapphire substrates. Data were taken under constant current condition. Maximum current levels were kept low to minimize Joule heating in the samples. Both four probe and two probe contacts gave the same results. All measurements reported here were done at zero magnetic field. 

\section{Results}
Figure 1 shows resistance-temperature ($R_o$-\textit{T}) plots for two SSCMO samples - one single crystal (with star) and another polycrystalline. The single crystal (SC) shows a first-order-like sharp MIT transition for $T \le T_{MI}$ whereas the transition to FM state in the polycrystalline (PC) sample is rather gradual. Such a behavior is believed to arise from a distribution of $T_{MI}$\cite {alexandrov06}. The hysteresis between cooling and warming cycles is also illustrative of first-order phase transition. Steady state \textit{I-V} measurements at $T \le T_{MI}$ were not feasible in the SC sample as the sharp transition is metastable with huge relaxation times ($>$ 5 hours). The same problem arises in PC samples during cooling cycle but is absent in the heating cycle. This makes PC samples a sort of necessity in the present study if measurements are to be carried out on both sides of the MIT. The \textit{$\Sigma$-V} characteristics of the SC in the warming cycle at different temperatures ranging from 81 to 95 K ($T> T_{MI}$) are plotted on a log-log scale in the left panel of Fig. 2a. The nonlinear response of the conductance to the application of bias \textit{V} is quite apparent in the figure. The onset bias $V_o(T)$ is the one at which conductance starts deviating from its linear value $\Sigma_o(T)$ at temperature \textit{T} and is defined (arbitrarily) such that $\Sigma(V_o) \approx 1.1\Sigma_o$. The plot of the normalized conductance $\Sigma(T,V)/ \Sigma(T,0)$ vs. the normalized bias $V/V_o$ using the data of the PI phase of Fig. 2a is shown in Fig. 2b which exhibits almost perfect data collapse, thus validating Eq. \ref {eq:scaling}. The bias scale $V_o$ appears to be independent of temperature and is shown in the inset of Fig. 2b (circles) as a function of the Ohmic conductance. The nonlinearity exponent $x^{PI}_T$ is thus zero in the PI phase of the SSCMO single crystal. The experimental path in this case corresponded to cooling the sample down to a temperature far below $T_{MI}$ and then gradual heating. To check the effect of possible hysteresis, the sample in another experiment was first cooled from the room temperature to 85 K slightly above $T_{MI}$ (i.e. still in the PI phase) where heating and cooling cycle coincided (Fig. 1), and then heated. The exponent $x^{PI}_T$ in this case turns out to be -0.31$\pm$0.06 (squares in the inset of Fig. 2b). Such sensitivity of the exponent to the experimental paths confirms the role of hysteresis.

\begin{figure}
\includegraphics[width=5.5cm]{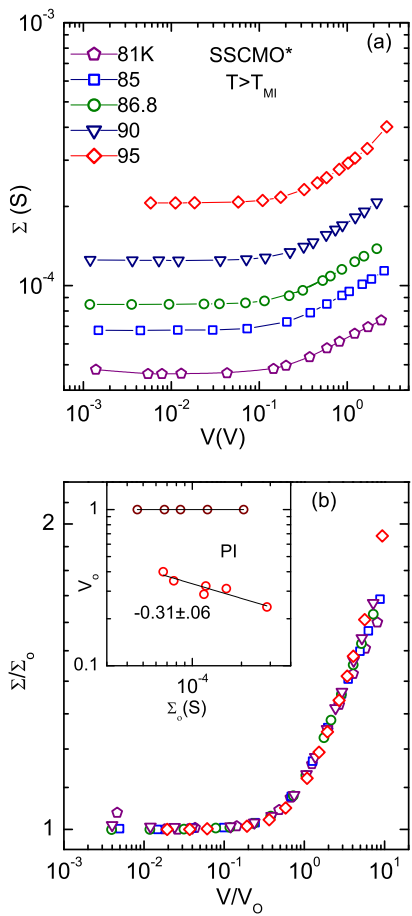}
\caption{(Color online) (a) Conductance vs. bias in the insulating paramagnetic (PI) phase of a SSCMO single crystal at different temperatures ($T>T_{MI}$) as indicated. (b) Scaling of the data in the panel \textit{a}. Inset shows bias scale, $V_o$ vs. linear conductance, $\Sigma_o$ for two experimental paths. See text for details. The number represents the slope of the linear fit to the data.}
\label{fig.2}
\end{figure}

\begin{figure}
\includegraphics[height=10cm]{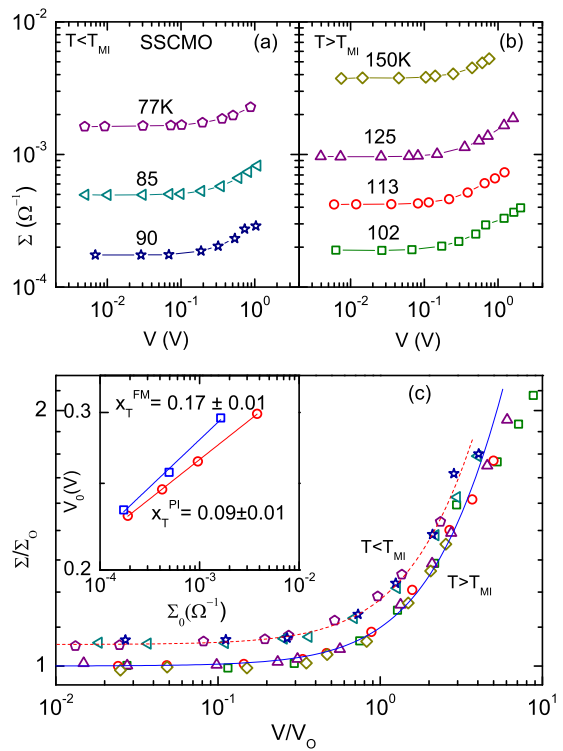}
\caption{(Color online) The upper panels show plots of conductance vs. bias in a SSCMO polycrystalline sample in the FM for $T<T_{MI}$ (a) and PI phase for $T>T_{MI}$ (b) at temperatures as indicated. The lower panel (c) shows the data collapse of the same data of the upper panels. The scaled curve of the FM side is shifted for clarity. Inset shows two log-log plots of $F_o$ vs. $\sigma_o$ data corresponding to PI (open circle) and FM sides (open square) respectively. The solid lines are linear fits to the data with slopes $x^{PI}_T$ and $x^{FM}_T$ respectively. }
\label{fig.3}
\end{figure}

Let us now turn to measurements in polycrystalline samples (Table I). Since results in various samples are basically similar, we consider only the results for the polycrystalline SSCMO shown in Fig. 3 for the purpose of comparing results from the single crystal of the same system discussed above. \textit{$\Sigma$-V} plots at various temperatures both below and above $T_{MI}$ are shown in Figs. 3a and 3b respectively such that in both cases conductance increases from bottom to top. Curves in panels \textit{a} and \textit{b} look qualitatively similar but exhibit a subtle difference as described below. Both sets of data were found to follow the same scaling as in Eq. \ref {eq:scaling} and actually collapse on the \textit{same} curve (Fig. 3c) (the scaled data in FM phase ($T<T_{MI}$) has been shifted in the figure for the sake of clarity). $V_o$ thus obtained are shown in log-log plots (open squares for the FM phase and open circles for the PI phase) in the inset of Fig. 3c. The power-law fits to Eq. \ref {eq:fscale} yield exponents on two sides of the MIT being different but positive and non-zero in contrast to zero value in the single crystal. Moreover, $F_o$ is generally less in polycrystals than in the single crystal (Fig. 4). This is contrary to what will be expected if the origin of nonlinearity were primarily due to intergrain boundary effects. $F_o$ as a function of conductivity for all samples are shown in Fig. 4. 

\begin{figure}
\includegraphics[width=5.8cm]{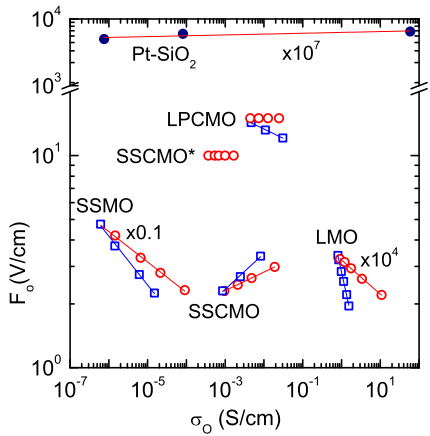}
\caption{(Color online) Plots of onset field $F_o$ vs. Ohmic conductivity $\sigma_o$ for different systems as marked. Circles and squares correspond to PI and FM phases respectively. Data of LMO and Pt-SiO$_2$ were shifted to right by the factors as indicated.}
\label{fig.4}
\end{figure}

Table I lists all the exponents extracted in this work. It may be noted that the nonlinear \textit{I-V} curves could also be caused by Joule heating\cite {mercone05b}. A simple mean field approach shows that $V_o \sim \Sigma_o^{-0.5}$ or, $x_{Joule} \approx -0.5$\cite {kkb99}. As none of the exponents obtained is close to this value it can be concluded that the nonOhmic behavior is an intrinsic feature and not caused by heating. If more current than used here is passed through the samples one would eventually encounter Joule regimes\cite {tokunaga04}.

\begin{table*}
\caption{\label{tab:table1}%
Sample parameters and nonlinearity exponents. See text for definitions and clarification on errors. The last two rows are obtained using data in Ref. \onlinecite {moshnyaga09}.}
\begin{ruledtabular}
\begin{tabular}{rccrrcc}

\textrm{System}&
\textrm{Abbre-}&
\textrm{Type} &
\textrm{$T_{MI}$} &
\textrm{$\Delta T_{1/2}$} &
\textrm{$x^{FM}_T$}&
\textrm{$x^{PI}_T$}  \\
&
\textrm{viation} &
&
\textrm{K} &
\textrm{K} & \\

\colrule
${\rm Sm}_{0.55}({\rm Sr}_{0.5}{\rm Ca} _{0.5})_{0.45}\rm{MnO}_3$  &SSCMO* & Single crystal &81 &  &  & 0~~~~   \\
${\rm Sm}_{0.55}{\rm Sr}_{0.3375}{\rm Ca}_{0.1125}{\rm MnO}_3$  &SSCMO & Polycrystal &94 &9 & $~~0.17 \pm 0.01$ & $~~~0.09 \pm 0.01$ \\
${\rm Sm}_{0.55}{\rm Sr}_{0.45}{\rm MnO}_3$	&SSMO & Polycrystal &69.5 &9.8 & $-0.23 \pm 0.01$ & $-0.14 \pm 0.01$  \\
${\rm La}_{0.275}{\rm Pr}_{0.35}{\rm Ca}_{0.375}{\rm MnO}_3$    &LPCMO & Polycrystal &113 &10.4 & $-0.09 \pm 0.01$  & 0~~~~ \\
${\rm La}_{0.87}({\rm Mn}_2{\rm O}_3)_{0.13}{\rm MnO}_3$ &LMO & Polycrystal &155 &93 & $-0.83 \pm 0.01$ & $-0.16 \pm 0.01$ \\
\hline
${\rm La}_{0.75}{\rm Ca}_{0.25}{\rm MnO}_3$    &LCMO (C) & Thin film &267 &10 & $~~0.27 \pm 0.04$  & $~~0.27 \pm 0.04$ \\
${\rm La}_{0.75}{\rm Ca}_{0.25}{\rm MnO}_3/{\rm BaTiO}_3$ &LCMO/BTO (CB) & Multilayer film &210 &30 & $-0.70 \pm 0.01$ & $-0.15 \pm 0.03$ \\
\end{tabular}
\end{ruledtabular}
\end{table*}

\section{Discussion}
Figs. 2, 3 and 4 unambiguously show that the field-dependent conductivities of various manganites indeed follow the same scaling (Eqs. \ref {eq:scaling} and \ref {eq:fscale}) as followed by VRH systems such as conducting polymers\cite {talukdar11} in spite of having very different microscopic pictures. This leads to the significant conclusion that the scaling of nonlinear conductivity like the phenomena of electronic localization is a general property of disordered systems and is independent of microscopic details. The scaling however is yet to be theoretically anchored. A critical phenomenon implies existence of a dominant length scale. It is not yet clear how such a length scale emerges out of the sea of disorder. The only model of disorder that is known to possess an intrinsic length scale (i.e. the correlation length, $\xi$) is that of percolation\cite {stauffer} (see below). Moreover, a two-component percolating system\cite {kkb97} does obey a scaling relation as in (\ref {eq:scaling}) for not too large current. The length scale in a thermodynamic system is basically determined by the distance from the critical temperature whereas the same in a transport process, according to Eq. (\ref {eq:fscale}), is determined solely by the ohmic conductivity or conductance. This means that the critical conductivity which is analogous to the critical temperature is zero. It may be recalled that the scaling theory of localization\cite {abrahams79} also assumes the Ohmic conductance as the scaling variable. Apart from the fact that a thermodynamic transition is an equilibrium process and a nonlinear transport process is an nonequilibrium process, there is an important difference between the former and the latter. The critical exponents belonging to a given universality class are fixed. But Table I reveals a plethora of exponent values belonging to the \textit{same} transport phenomena (i.e. MIT) in different manganite systems. A variety of exponent values was also observed in case of conducting polymers\cite {talukdar11}. In critical phenomena there are quantities such as the proportionality constant in (\ref {eq:fscale}) or the critical amplitude, which are essentially determined by the system at hand. This of course requires an understanding of the microscopic picture. Presently, a complete theory for nonlinear transport in the vicinity of MIT in manganites is lacking. The exponents and the field scales are further discussed below. 

The most frequently discussed model used to describe transport in various regimes of manganites, including that of the MIT, invokes the idea of percolation\cite {pinaki09,yunoki98,dagottobook}. Coexistence of domains of metallic FM phase and insulating PI phase\cite {uehara99} with a temperature-dependent metallic fraction \textit{p}\cite {alexandrov06} apparently finds a ready analogy in the standard (i.e. classical) percolation model\cite {stauffer}. As the temperature $T>T_{MI}$ is decreased, the FM phase starts growing and ultimately reaches the percolation threshold at $T=T_{MI}$. Gefen et al. \cite{gefen86} have suggested two theoretical models for onset of nonlinear conduction in a binary mixture. One (NLRRN) assumes each conducting bond to be nonlinear possibly due to joule heating and predicts the exponent to be given by $x_p \approx -0.03(0.03)$ in 3D(2D)\cite {aharony87}. This is incompatible with the experiments with real binary systems\cite{gefen86,kkb97}. The another (DRRN) ascribes the onset of additional conduction due to hopping or tunneling across insulating bonds under sufficiently strong field. In this case the onset bias is given by $V_o \sim \xi^{-1}$ and hence, $x_p \le \nu / t \approx 0.45(1)$ in 3D(2D) where $\nu$ and \textit{t} are the correlation and conductivity exponents, respectively. This value is consistent with the experimental values \cite{gefen86,kkb97}. As already noted\cite {pinaki09}, several fundamental disagreement between manganites and standard percolation theory seem to exist - i) the conductivity exponents as high as about 7 required to fit data in manganites are much higher than that (~1.9 in 3D) in the standard percolation model; ii) the MIT can be first order in manganites in contrast to being second order in the percolation model. This work on nonlinear conduction now throws up another couple of disagreements: iii) the scaling function in the PI phase does not match the one in a composite below the percolation threshold; iv) instead of a single value of an exponent in a thermodynamic transition, a multitude of values (Table I) of the nonlinearity exponent is found. Even then, it is seen that none of the exponents in FM phase is compatible with the value of 0.45.

Mosgnyaga et al.\cite {moshnyaga09} proposed a specific mechanism involving correlated polarons for nonlinear transport around the MIT in manganite samples. The resistance behavior with temperature in the PI phase can be accounted by the correlated polarons. The authors argued that the electric nonlinearity resulted from the coupling of these polarons to the electric fields. The sample resistance is expressed as
\begin{equation}
R \sim \frac{1}{N - N_{CP} +AN_{CP} I^2}
\label{eq:RI}
\end{equation}
where $N_{CP}$ and $N \sim 1/R_o$ are the concentrations of correlated polarons and charge carriers respectively and \textit{A} is a constant. Note that the above equation upon incorporation of current is a more consistent form of the equation originally considered by the authors. Assuming $N_{CP} \ll N$, one gets $R = R_o -R_{3w} $ where
\begin{equation}
R_{3w} \sim R^2_o N_{CP} I^2
\label{eq:RNcp}
\end{equation}
$R_{3w} $, also known as the third harmonic resistance, can be related to the scaling in (\ref {eq:scaling}) as shown in the next section where a new method of extracting the nonlinearity exponent is also discussed.

\begin{figure}
\includegraphics[width=6cm]{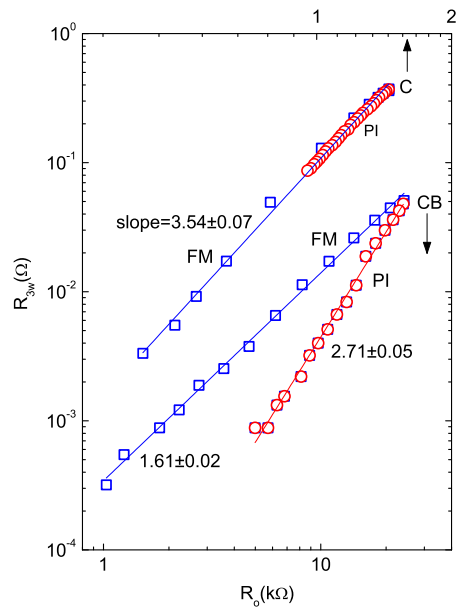}
\caption{(Color online) Plots of the third harmonic resistance $R_{3w}$ vs. the Ohmic resistance $R_o$ in two systems as marked. Data are taken from Fig. 3 of Ref. \onlinecite {moshnyaga09}. Circles and squares correspond to PI and FM phases respectively. Slopes are as indicated.}
\label{fig.5}
\end{figure}

\subsection{Third harmonic resistance, $R_{3w}$ }
The third harmonic method is based upon the assumption that voltage \textit{V} of a nonlinear \textit{I-V} curve can be written in odd powers of current:
\begin{equation}
V = R_o I + {\alpha}_3 I^3 +...
\label{eq:VI1}
\end{equation}
where $R_o=1/ \Sigma_o$ is the linear resistance and ${\alpha}_3$ is a coefficient. For an ac $I=I_a \sin wt$ with the fundamental frequency $w$, the first nonlinear term on the right hand side of the above equation gives rise to a third harmonic term $V_{3w}(w)=V_{3w} \sin 3wt$ with $V_{3w}={1 \over 4} {\alpha}_3 {I_a}^3$. The third harmonic resistance is simply defined as $R_{3w}= V_{3w}/I_a = {1 \over 4} {\alpha}_3 {I_a}^2$. The nonlinear resistance \textit{R} can be written as $R(I_a) = 1/ \Sigma = R_o + {1 \over 4} {\alpha}_3 {I_a}^2$. A comparison with the scaling equation (\ref {eq:scaling}) allows us to express the coefficient ${\alpha}_3$ in terms of the scaling quantities. For this purpose it is convenient to rewrite Eq. \ref {eq:scaling} in terms of \textit{R} and \textit{I}:

\begin{equation}
{R(M,I) \over R_o} = {\Phi}_I \left ({I \over I_o} \right ).
\label{eq:Rscaling}
\end{equation}
Here $\Phi_I$ is a new scaling function and 
\begin{equation}
I_o (M) = \Sigma_o V_o \sim {\Sigma_o}^{1+x_M}.
\label{eq:iscale}
\end{equation}

Expansion of $\Phi_I$ in even powers of \textit{I} yields $\Phi_I = 1 + b(I/I_o)^2+...$ where \textit{b}, in contrast to ${\alpha}_3$, is a constant independent of $R_o$. Using (\ref {eq:iscale}) we obtain $R(M,I) = R_o + b' {R_o}^{3+2x_M} I^2$ where \textit{b'} is a constant. Comparison of the two expressions for \textit{R} yields $R_{3w}$ proportional to a power of $R_o$ with the exponent $3+2x_M$ at a fixed current:
\begin{equation}
R_{3w}(M) \sim R_o ^{3+2x_M} I^2.
\label{eq:r3w}
\end{equation}
Note that this relation between $R_{3w}$ and $R_o$ is a perfectly general one, following directly from the scaling (\ref {eq:Rscaling}) and (\ref {eq:iscale}) and the assumption of analyticity of the scaling function. It is seen that Eq. \ref {eq:r3w} provides for an alternative means of determining the nonlinearity exponent $x_M$ from the slope of log-log plot of $R_{3w}$ vs. $R_o$. Moshnyaga et al.\cite {moshnyaga09} have measured these resistances around the MIT in several epitaxial thin films of ${\rm La}_{0.75}{\rm Ca}_{0.25}{\rm MnO}_3$ as functions of temperature at a fixed current. Data from two such samples labeled C and CB (see Fig. 3 of Ref. \onlinecite {moshnyaga09} ) are replotted in Fig. 5 using log-log axes. The linearity of the plots indicates strong validity of Eq. \ref {eq:r3w} and consequently, scaling assumption (\ref {eq:iscale}). The exponents $x_T$ derived from the slopes are shown in Table I. It would have been worthwhile to compare the exponents thus obtained with ones determined from scaling of \textit{I-V}s as described earlier. It follows from (\ref {eq:r3w}) and (\ref {eq:RNcp}) that $N_{CP} \sim R^{1+2x_M}_o$. It would be interesting to obtain an independent verification of this prediction possibly from the neutron scattering data since the scattering intensity is supposed to be proportional to $N_{CP}$.

According to Eq. \ref {eq:r3w}, the frequency dependence of $R_{3w}$ is entirely determined by that of $R_o$. Since $R_o(w)$ in disordered regimes usually decreases with frequency\cite {chaudhuri07,*cbb91}, $R_{3w}$ is also predicted to decrease with frequency. However, this seems to be in contradiction with the data in Fig. 5 of Ref. \onlinecite {moshnyaga09} where $R_{3w}$ increased with increasing frequency. Note that owing to the one-parameter scaling the nonlinear resistance $R_{pw}$ corresponding to the term in (\ref {eq:VI1}) with a power \textit{p} can be expressed in terms of $R_o$:
\begin{equation}
R_{pw}(M) \sim R_o(M) ^{p+(p-1)x_M} I^{p-1}.
\label{eq:rpw}
\end{equation}
\textit{p} is tacitly assumed to be an integer. However, the above relation still remains valid even when the powers may not be integer. Possibility of such situations are discussed next.

\subsection{Scaling function, $\Phi$}
One of the experimental features of the scaling functions in manganites is that the maximum value of the normalized conductance (Figs. 2b and 3c) hardly exceeded the factor of two compared to several orders of magnitude in conducting polymers \cite{talukdar11}. It makes drawing of any conclusion about its asymptotic nature at large field rather uncertain. In order to ascertain the nature of the scaling functions, scaled data belonging to the PI phase of the single crystalline and polycrystalline SSCMO are superposed in Fig. 6. The scaling functions are seen to diverge at higher fields. This may be due to grain boundary effects in polycrystalline samples. Such details along with effect of hysteresis on exponents illustrate well capabilities of transport measurements in nonOhmic regimes as useful probes. Note that as explained before, the data curves as well as other curves are drawn such that each curve passes through the point (1,1.1) indicated as the intersection of two perpendicular lines. The functional forms for various possible nonlinear effects in mangenites including one based upon multistep tunneling (GM)\cite {matveev88} have been discussed in literatures\cite {paranjape03}. Accordingly, a functional $\Phi (z) = 1 +0.1 z^q$ is tried ($z=V/V_o$). Three curves corresponding to  \textit{q}= 0.5 (dash), 1 (dot) and 4/3 (dash-dot) are shown in Fig. 6. It is seen that the term $z^{4/3}$ rises much faster than the data in single crystal as well diverging from the data in the polycrystalline sample although less rapidly. The GM model has been very successful in fitting the nonlinear data in conducting polymers up to several orders of magnitude\cite {talukdar11}. But this model seems quite inappropriate to describe the nonlinear data in manganites even though the increase in conductance in this case is much smaller than in the conducting polymers. In fact, it should be of no surprise that the GM model fails since no basis for application of the model exists particularly in the single crystal. Clearly, the first power in a series for $\Phi$ must be less than 4/3. Thus, the scaling functions obtained in this work are not analytic and are therefore incompatible with the expansion like Eq. \ref {eq:VI1}. The curve with \textit{q}=1 appears to cross the single crystal data at about $z \sim 1.7$, varying slower at lower values and somewhat faster at higher values. The term with \textit{q}=0.5 is incompatible with the data at hand.

\begin{figure}
\includegraphics[width=6.5cm]{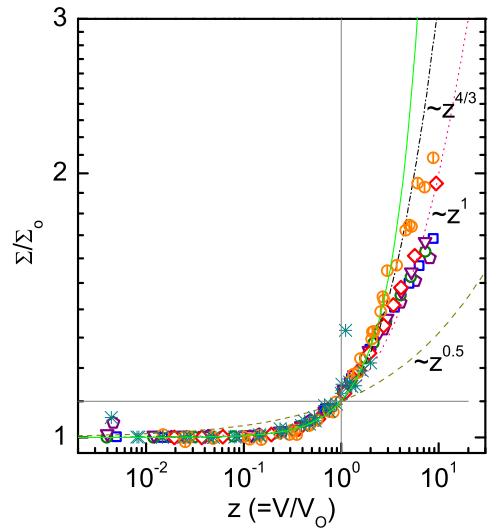}
\caption{(Color online) Scaled data in PI phase of SSCMO single crystal of Fig. 2 (noncentered symbols) and polycrystals of Fig. 4 (circles) are shown for comparison. The lines (except the solid one) are plots of $1+0.1 z^q$ with \textit{q}=0.5 (dash), 1 (dot) and 4/3 (dash-dot). The solid line is a fit to $1+0.1 z^{4/3}+0.0014 z^{18/5}$ of the scaled data of Pt-SiO$_2$. See text for details.}
\label{fig.6}
\end{figure}

From the viewpoint of percolation theory the PI phase is \textit{below} the threshold since there is no FM cluster that spans the entire sample. A PI phase at different temperature is comparable to a non-percolating granular samples with different conducting fractions (\textit{p}). We have analyzed such data at a fixed temperature of three samples of Pt-SiO$_2$ with different metal fractions available in the literatures\cite {abeles75}. The scaled data with $\Sigma / \Sigma_o$ being as high as 30 can be fitted very well to a GM expression: $1+0.1 z^{4/3}+0.0014 z^{18/5}$ (solid curve in Fig. 6). Comparison with the manganite data shows that the nonOhmic conductivity in the latter increases much slower than that in a binary composite. Interestingly, the observation of the same scaling function above and below $T_{MI}$ in the polycrystalline samples (Fig. 3) is not expected from the simple percolation theory. In the FM phase which corresponds to a percolating system \textit{above} the percolation threshold, the nonlinear conduction  is thought to occur due to opening of new conduction channels across thin insulating bridges connected to the backbone\cite {gefen86,kkb97}. But in the PI phase an electron must tunnel from one metallic cluster to another across insulating barriers in absence of any backbone. On the other hand, within the scenario that assumes an existence of a distribution of $T_{MI}$\cite {alexandrov06} in the PI phase, the presence of a fraction of PI clusters may give rise to same scaling functions on both sides of $T_{MI}$ as observed.

\subsection{Nonlinearity exponents, $x_T$ }
At this stage due to lack of a proper theory, the exponents can not be related to the microscopic features. Nevertheless, its sheer variety in signs and numbers in Table I reinforces the potential utility of the nonOhmic probe. We note the followings:
\begin{enumerate}
 \addtolength{\itemsep}{-0.65\baselineskip}
 \item The exponent $x_p$ in a non-percolating systems such as Pt-SiO$_2$ (Fig. 4) is found to be nearly zero and hence, consistent with  $x^{PI}_T$ in SSCMO*.
 \item The signs as well as the magnitudes of $ x^{FM}_T $ are generally different from $\sim$0.45 found in the three-dimensional binary composites above the threshold\cite {kkb97}. The NLRRN model predicts a negative exponent but its magnitude is difficult to compare with the array of experimental numbers.
 \item As observed in conducting polymers, the exponent value may depend upon the experimental path in the variable space. In the present case the path may be changed due to hysteresis (Fig. 2b).
 \item $x^{PI}_T$ can be both positive and negative.
 \item $x^{PI}_T$ has the same sign as corresponding $x^{FM}_T $.
 \item $|x^{PI}_T|$ is generally less than $|x^{FM}_T|$ (the exponents may be occasionally equal (within error) as in the sample C).
 \item The errors in exponents in Table I only refer to those of least-square fittings. It does not include uncertainties in actual scaling procedures and/or digitization of data (e.g. from Ref. \onlinecite {moshnyaga09} in the present case), whenever applicable. Such uncertainties can increase errors by another 10-15$\%$ depending upon the quality of data under consideration.
\end{enumerate}
The different exponents on two sides of the transition in polycrystalline samples must be related to the subtle differences in microstructures. It has been recognized that the finite width of the \textit{R-T} curve at $T< T_{MI}$ is due to a distribution of $T_{MI}$'s belonging to various domains of PI subjected to different degree of disorder\cite {alexandrov06}. Let us define the width by $\Delta T_{1/2}=T_{MI}-T_{1/2}$ as the difference between $T_{MI}$ and the temperature $T_{1/2}$ at which the resistance is half of the peak resistance i.e. $R_o(T_{1/2})= R_o( T_{MI}) /2$ (see inset of Fig. 7). In case of a single crystal, $\Delta T_{1/2}$ will be close to zero. The absolute difference in exponents, $\Delta x_T = |x^{PI}_T - x^{FM}_T|$ is plotted against $\Delta T_{1/2}$ in Fig. 7 and is seen to be an increasing function of the width. The dashed line is an empirical fit of the data to $\Delta x_T = 0.7{1- \exp [-0.07(\Delta T_{1/2} -\Delta T_o) ] } \text{ for } \Delta T_{1/2} \ge \Delta T_o$, and $0\text{ for } \Delta T_{1/2} \le \Delta T_o$ where $\Delta T_o \approx $ 8K. Sharper is the resistance fall on FM side, closer is $|x^{FM}_T|$ to $|x^{PI}_T|$. It may be noted that the width $\Delta T_{1/2}$ is ultimately related to the width of the distribution of $T_{MI}$.

\begin{figure}
\includegraphics[width=5cm]{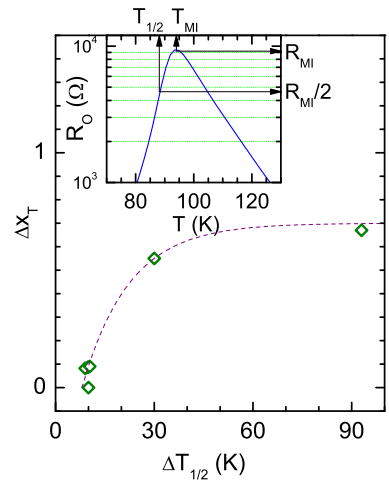}
\caption{(Color online) Plot of $\Delta x_T~(= |x^{PI}_T - x^{FM}_T|)$ vs. $\Delta T_{1/2}~(=T_{MI}-T_{1/2})$ for six samples in Table I. The dashed line is essentially a guide to the eye. The inset illustrates using SSCMO data in Fig. 1 how the quantities in $\Delta T_{1/2}$ are defined. See text for details.}
\label{fig.7}
\end{figure}
 
\subsection{Field scale, $F_o$ }
Field scales determined according to the criterion assumed in this work ($\sigma(F_o) = 1.1\sigma_o$) are displayed in Fig. 4 for all samples including SSCMO* and Pt-SiO$_2$. It is seen that the scale in SSCMO* is about three orders of magnitude less than that in Pt-SiO$_2$. This can be understood by noting that a carrier moves from one metallic cluster to another which are chemically and structurally same but distinguished only by spin or charge ordering. Hence, the energy barrier that a carrier must overcome must be much less than that in a medium of chemically dissimilar materials like Pt and SiO$_2$. On the other hand, taking the separation distance in a manganite as large as 100 nm\cite {uehara99} a field of 10 V/cm ($F_o$ in SSCMO*, Fig. 3) corresponds to an anomalously low barrier height of about 1 K compared to about 300 K estimated from noise studies\cite {raquet00a}. This may corroborate the claim of electronically soft phases in manganites\cite {milward05}.

The issue of the field scale is really related to the issue of existence of an intrinsic length scale in a disordered medium as argued in the beginning of Section IV. However, its magnitude like the nonlinearity exponent must be determined by the microscopic details of the system at hand. A proper theoretical framework is still to come. In fact, the straightforward application of the prevalent theories\cite {abeles75, shklov76} using one-electron picture leads to inconsistent conclusions.  The field scale $F_o$ is given by the condition that the field-activated conduction becomes comparable to the temperature-activated one i.e.,
\begin{equation}
eF_o w = k_B T.
\label{eq:fscaleT}
\end{equation}
Here \textit{w}, a relevant length scale, may be the mean separation length between two metallic clusters in a granular system \cite {abeles75} or the average hopping distance in a VRH system\cite {shklov76}. The scale $F_o$ is decided by the behavior of \textit{w(T)}. In view of different values of the exponents $x_T$, it is difficult to see how a single physical parameter can account this. It remains to be seen that how introduction of correlation among electrons as in manganites will affect a relation like Eq. \ref {eq:fscaleT}. It may be noted that  the latter is not compatible with  Eq. \ref {eq:fscale} which is a typical power-law relation in critical phenomena.

\section{Conclusion}
In summary, this work on scaling of nonOhmic conductivity in manganites highlights the fact that an intrinsic length scale, just as localization, is a general property of a disordered medium. The length scale leads to the associated scaling property. At the same time, this work also shows how studies of the scaling quantities which depend on the microscopic details could be an useful probe of complex systems such as manganites. The phenomenology of the scaling now needs to be supported by a proper theoretical analysis to explain, among other things, the great variety of exponents found in this work - a novel feature that distinguishes transport transitions from thermodynamic ones. It is hoped that this work would encourage systematic study of other disordered regimes such as charge ordered states from the perspective of scaling.

\section{Acknowledgments}
This work was supported in part by DST, India through Project No: SR/S2/CMP-0054/2008. Authors wish to thank S. Giri for the LMO sample, A. De for the cryotip facility and S. Mukhopadhyay for useful discussion.

\bibliography{nonohmic_29}

\end{document}